# DCT-domain Deep Convolutional Neural Networks for Multiple JPEG Compression Classification

Vinay Verma, Nikita Agarwal, and Nitin Khanna[*]

*Multimedia Analysis and Security (MANAS) Lab,*
*Electrical Engineering, Indian Institute of Technology Gandhinagar (IITGN),*
*Gujarat, India*

## Abstract

With the rapid advancements in digital imaging systems and networking, low-cost hand-held image capture devices equipped with network connectivity are becoming ubiquitous. This ease of digital image capture and sharing is also accompanied by widespread usage of user-friendly image editing software. Thus, we are in an era where digital images can be very easily used for the massive spread of false information and their integrity need to be seriously questioned. Application of multiple lossy compressions on images is an essential part of any image editing pipeline involving lossy compressed images. This paper aims to address the problem of classifying images based on the number of JPEG compressions they have undergone, by utilizing deep convolutional neural networks in DCT domain. The proposed system incorporates a well designed pre-processing step before feeding the image data to CNN to capture essential characteristics of compression artifacts and make the system image content independent. Detailed experiments are performed to optimize different aspects of the system, such as depth of CNN, number of DCT frequencies, and execution time. Results on the standard UCID dataset demonstrate that the proposed system outperforms existing systems for multiple JPEG compression detection and is capable of classifying more number of re-compression cycles then existing systems.

## 1. Introduction

Digital images are ubiquitous due to the advances in imaging technologies and low-cost availability of imaging devices like handheld digital still cameras and mobile phones. Most of these hand-held devices such as mobile phones come equipped with network connectivity and provisions for uploading these images on different social media platforms. An enormous number of digital images are generated at a rapid rate every day. These images can be easily enhanced for better visualization, and can also be manipulated to change the original information contained within them, thereby their meaning. Even a non-expert can perform several image processing operations such as retouching, rescaling, and cropping due to the wide availability of image processing software, e.g., Photoshop, Gimp, Picasa, PicsArt, etc. to name a few. Many of the manipulations can be done without leaving any visual traces or artifacts of tampering in the manipulated images. These images can be easily shared and distributed using social media platforms such as Instagram, Twitter, and Facebook which often leads to spreading of wrong information and severe consequences resulting from it.

Digital images can also act as substantial pieces of evidence in the court of law, where the authenticity and integrity of the image is an utmost priority and critical [1, 2]. Researchers have made several attempts to address the issue of integrity and trustworthiness of digital images without having any prior information about the concerned image [3, 4]. In literature, there are studies with the specific focus on solving the issue of image authenticity and integrity by use of compression based forensics, in particular, JPEG-based forensics. Finding out the compression history of image addresses the forensic problem of establishing the integrity and trustworthiness of the image. JPEG compression based forensics are popular because most of the cameras encode the acquired images in the JPEG format to save onboard memory and most of the images available on the Internet are also encoded in JPEG format. The image under inspection may be directly coming from the source camera, or it may be decompressed, manipulated and recompressed multiple times in JPEG format.

Multiple compression can occur when an image has gone through a chain of compression and decompression steps. For example, in a practical scenario, the image captured from most of the cameras is encoded in the JPEG format by the camera itself due to storage limitations which is the first compression in the series. Then the same image may be decompressed, enhanced, manipulated, and re-saved in the JPEG format resulting in double compression. The number of compressions keeps in-

---

[*]Corresponding author. Complete code of the proposed system will be made publicly available along with the published version of the paper. Presently, the code is submitted as additional material and is made available to the reviewers.
*Email address:*
`{vinay.verma,nikita.agarwal,nitinkhanna}@iitgn.ac.in` (Vinay Verma, Nikita Agarwal, and Nitin Khanna)

creasing every time image undergoes any of such modification, and re-saved in the JPEG format. And enabling forensic experts to use this information to identify the probable manipulation of the image

There are studies which focus on the problem of double JPEG compression detection. If the image is single compressed, it can be claimed that the image under inspection is directly coming from the camera, without any intermediate manipulation. On the other hand, if the image is detected as double compressed, it can be established that image has been at least once opened in some image processing software and further saved in JPEG format. However, nowadays due to the involvement of social media platforms in sharing most of the multimedia data, in particular, digital images, even the authentic image can have the traces of double compression in it. The second compression is applied when the image is uploaded to the social media platform such as Whatsapp Messenger, Facebook, Twitter, Instagram, etc. This makes the double compression detection algorithm fallible and indicates that there is a need for reliable methods of multiple JPEG compression detections. Detecting multiple compression histories can enable us to verify the claim of authenticity and trustworthiness of the images posted on social media platforms. The proposed method can also be used in steganalysis [5] and forgery identification [6]. The main contributions of this paper are the following:

- Use of CNN for multiple JPEG compression detection while CNN based system existing in the literature only targeted double compression detection (Section 2)

- Design of appropriate pre-processing stage before feeding the image data to CNN, which directly utilizes JPEG bit stream and reduces the content dependent nature of the data fed to the CNN (Section 3)

- Robust performance even at a larger number of compression stages (Table 10, Section 4.6)

- Average classification accuracy of 91% for classifying patches of size $128 \times 128$, thus paving a way for possible application in forgery detection (Section 4.8)

Rest of the paper is organized as follows. In Section 2, previous work on JPEG compression using handcrafted features as well as CNN is described. Section 3 contains the proposed method. Experimental results are described in Section 4, while conclusion and future work is discussed in Section 5.

## 2. Related Works

CNNs have been used in many speech, image and video classification and recognition tasks. Many multimedia forensics related tasks such as median filtering detection [7], camera model identification [8, 9, 10], forgery detection [11], and steganalysis [12, 13, 14, 15] also has been addressed with the aid of CNNs. This section presents a brief summary of systems proposed in literature for compression based forensics, either using handcrafted features or data-driven CNN based systems.

The problem of multiple compression detection is a relatively new research area. But the detection of double JPEG compression in images has been previously explored in [16, 17, 18, 19, 20, 21] using handcrafted features. Due to current advancements in deep learning, there are attempts made by the forensic community to solve the problem of double JPEG compression detection using data-driven approach. Some recent works [22, 23, 24, 25] on double JPEG compression detection uses Convolution Neural Network (CNN) based approaches.

### 2.1. Compression Detection using Handcrafted Features

The problem of single vs. double compression detection and primary quantization matrix estimation has been explored in [16], in which authors, used normalized histogram of the JPEG coefficients (that is quantized Discrete Cosine Transform (DCT) coefficients) to detect the single vs. double compression, by analyzing artifacts like missing coefficients and double peaks in the histogram. Authors in [17] studied that the double compressed images exhibited a periodic artifacts in the histogram of JPEG coefficients and computed Fourier transform of the resulting histogram to detect the traces of double compression. The performance of algorithm is evaluted with 100 natural images. Fu *et al.* [18], established the generalized Benford's law for the JPEG coefficients. Authors demonstrated that probability distribution of first digits in all sub-bands(excluding DC (average) coefficient) from each $8 \times 8$ non-overlapping blocks of JPEG coefficients of a single compressed image follow the generalized Benford's law, while the image undergone double compression deviates from the law. This study does not report the experimental results for single vs. double compression detection. Extending the idea established in [18], Li *et al.* [19], designed 180-D feature vector from first 20 AC sub-bands(in zig-zag order) of $8 \times 8$ non overlapping blocks of the JPEG coefficients by calculating the probability distribution of first significant digits (1-9) from each of the 20 sub-bands. These feature vectors are named as mode-based first digit features (MBFDF), and fed to the supervised classifier to classify between single vs. double compressed images. For the evaluation of results, 1138 images from UCID (Uncompressed Colour Image Dataset) [26] dataset are used for training the Fisher Linear Discriminant(FLD) classifier and remaining 200 images are used for testing purpose. Amerini *et al.* [20] also used similar kind of features for splicing localization, but with the difference that only first nine sub-bands, excluding the DC coefficients, from each $8 \times 8$ blocks of JPEG coefficients are used for calculating first digit histogram. In final feature selection, only the occurrence of digit (2,5 and 7) is chosen, resulting in $9 \times 3 = 27$ dimensional feature vector for each image. 1338 images of the UCID [26] database are used for training the Support Vector Machine(SVM) classifier and 1448 images of Dresden Image Database [27] is used for testing purpose. Taimori *et al.* [21] calculated feature vector for each image based on the Benford's law. Given an image DCT coefficients are extracted and all the 63 sub-bands(excluding the DC coefficient) are chosen from each $8 \times 8$ block, probability mass functions of first significant digits(1-9) and the probability of digit 0 from the second significant digit are estimated. This results in 630-dimensional(63*(9+1)) feature vector for each image. Three learning strategies that are bottom-up, top-down and combination of these two have been proposed for double compression detection, forgery localization and estimation of the first quantization table.

Detection of multiple JPEG compression has been addressed in [28, 29]. Chu *et al.* [30] established that using features from normalized DCT coefficient histogram of sub-band (2,3), maximum four number of compressions can be detected. Similar kind of theoretic limit on the number of compressions can be established using other features used in JPEG compression detection. Pasquini *et al.* [28] proposed a method to detect multiple numbers of JPEG compression based on the Benford-Fourier coefficients. Authors have reported the results up to three JPEG compression detection. Milani *et al.* [29], have also given a method to detect the multiple JPEG compressions. They have considered the detection of images compressed up to four times, and have also shown some result for the images compressed up to five times. Authors have used first nine spatial frequencies, and for first significant digit histogram, probability mass function of only digits {2,5,6} is used, resulting in final feature dimension of twenty-seven. For the evaluation of result, 100 images from UCID dataset are used for training the SVM classifier and 200 images are used for testing purpose. The average accuracy of close to 88% is reported for detecting up to four number of JPEG compression with images compressed with the last quality factor of 80.

*2.2. Compression Detection using CNN*

Wang *et al.* [22], proposed the detection of double JPEG compression based on CNN. For the input to CNN, they used the histogram of first nine sub-bands (excluding DC) in zig-zag order from the blocks of $8 \times 8$ DCT coefficients for each image. For each sub-band histogram, bin location is restricted only to the values from {-5,-4,-3,-2,-1,0,1,2,3,4,5}. So for each of the nine sub-bands, an eleven-dimensional feature vector is obtained resulting in a 99-dimensional feature vector for the input to CNN. Uricchio *et al.* [23], have proposed the multi-domain CNN for discriminating among uncompressed vs. single compressed vs. double compressed, in which three is a 'spatial-domain CNN' which takes input as three channel color patch, and in another 'frequency-based CNN' input to the model is a histogram of DCT coefficient. For calculating the histogram of the DCT coefficients, they considered the DCT coefficients in the range {-50,-49,...,49,50}, from the first nine sub-bands (excluding DC) in zigzag order. That results in $101 \times 9 = 909$ dimensional feature vector. Both the CNN models are tested individually and were also combined to results in 'multi-domain CNN. Combined CNN model performed better as compared to the individual CNN. Barni *et al.* [24] has used CNN for aligned as well as non-aligned double JPEG compression detection by using input as image patches of size $64 \times 64$ and $256 \times 256$ to the CNN. RAISE [31] dataset is used for evaluating the performance of the algorithm. Li *et al.* [25] addressed the problem of detecting the double compression using multi-branch CNN. Moreover, they have used raw DCT coefficients of first 20 AC sub-bands from each $8 \times 8$ block in zigzag order; each sub-band is fed to one branch of the CNN. Moreover, in one branch whole 20 sub-bands are fed as a tensor having the third dimension of 20. Resulting 21 CNN branches are combined to produce a multi-branch CNN architecture.

To the best of our knowledge, there is no existing work that uses data-driven learning capability of convolution neural network for addressing the problem of multiple JPEG compression detections.

## 3. Proposed Model

The system proposed in this paper aims to differentiate images based on the number of compressions they have undergone, independent of their scene content. Recently CNNs have been used with great success in a number of tasks related to content-based image retrieval, using architectures such as AlexNet [32], VGG Net [33], GoogLeNet [34] and ResNet [35]. Although, these CNNs used networks with different architecture and varying depth, the input to the networks are always images in the spatial domain. In contrast, in the forensic problem addressed in this paper, we are essentially aiming at differentiating images based on traces of quantizations, rounding and truncation noise present in them, independent of the image content. Thus, instead of directly feeding images in spatial-domain or pixel-domain into CNN, the proposed system first extracts suitable features from these images and then feeds these features into a CNN of appropriate architecture and depth.

Figure 1 shows an overview of the proposed system, whose input is an image $I$ in JPEG format and output is class label $\hat{L}$ corresponding to the number of compressions that image has undergone. During the training phase, the correct number of compressions $L$ undergone by the image are also inputted to the system as ground truth.

*3.1. Feature Extraction*

Given an image, JPEG compression involves performing DCT independently on each of the $8 \times 8$ block of the image. Each $8 \times 8$ block of DCT coefficients are quantized with JPEG quantization tables which differ for Luminance and Chrominance channels. These quantized DCT coefficients are termed as "quantized DCT coefficients" or "JPEG coefficients". Entropy encoding is performed on quantized DCT coefficients to obtain the JPEG bitstream. To get the image back in the spatial domain, operations in reverse order such as entropy decoding, dequantization, inverse DCT (IDCT) and finally rounding and truncation is performed. This inverse chain of operations to get the image back in pixel domain is referred as decompression. In this paper, we have directly extracted the quantized DCT coefficient from the JPEG bitstream, instead of first decompressing the image and then finding the quantized DCT coefficient.

Histograms have been successfully used in a number of tasks related to images where global information from an image need to be captured without focussing on local structures present in an image. Since the quantization, rounding and truncation operations happening in JPEG compression apply the same procedures independent of the image content, usage of histogram-based features is apt for this problem. Further, quantization in JPEG compression is applied in DCT domain, and quantization step size depends on the location of DCT frequency and not on the spatial location or content of a particular 8x8 block. Thus, independent of the spatial location of a particular 8x8 block, a particular DCT frequency corresponding to it, say $(k_1, k_2)$ where $k_1, k_2 \in 1, 2, \ldots, 8$, meets the same treatment as frequency $(k_1, k_2)$ in any other block. Hence, the proposed feature extraction step utilizes histograms of different DCT frequencies. Further, since the input to the proposed system is always a JPEG image, unlike some of the existing systems which first read the image's pixel values (effectively doing decompression) and then perform DCT on these pixel values to obtain DCT coefficients corresponding to an input image, the proposed system reads in the raw DCT values directly from the JPEG bitstream. We have used an opensource python package, pysteg [36] for this purpose.

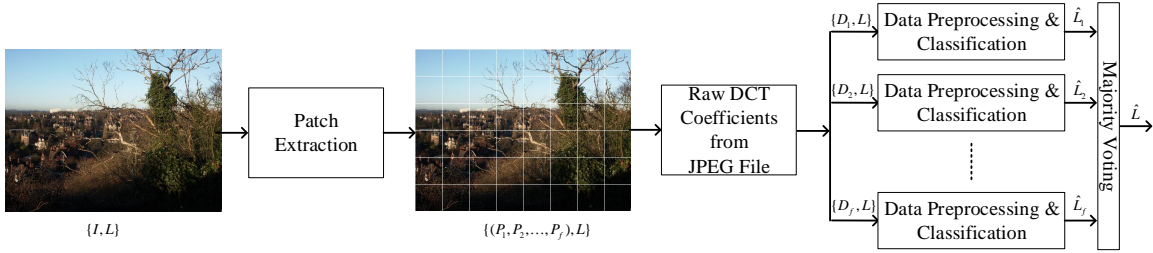

Figure 1: Overview of the Proposed System

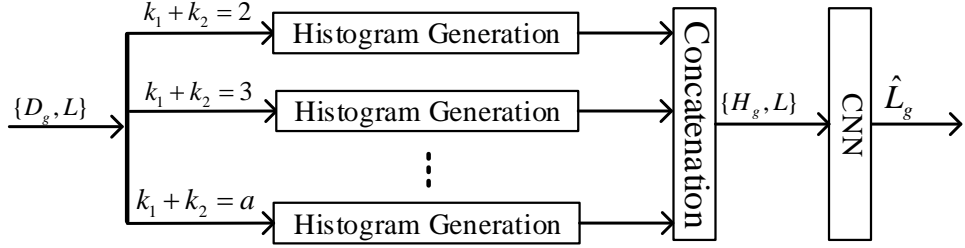

Figure 2: Data Preprocessing and Classification

The first step of the system is to divide the complete input image into smaller patches of size $N_p \times N_p$ (block labeled as "Patch Extraction" in Figure 1). For an image of size $M_1 \times M_2$, this will result in $f$ number of patches, $\{P_1, P_2, \ldots, P_f\}$, where $f = \lfloor (M_1/N_p) \rfloor * \lfloor (M_2/N_p) \rfloor$. Each of these patches are independently processed and classified using CNN and finally their decisions, $\{\hat{L}_1, \hat{L}_2, \ldots, \hat{L}_f\}$, are merged using majority voting to decide the final predicted class of the image, $\hat{L}$. If the image dimensions are not a multiple of $N_p$, then data from some of the rows/columns are discarded. Size of the patches $N_p$ is kept a multiple of 8 as the JPEG compression algorithm independently performs quantization on blocks of size $8 \times 8$. $N_p$ should be large enough so that each of the patches will have a sufficiently large number of $8 \times 8$ blocks, giving us statistically significant histograms of DCT frequencies. At the same time, we want to have a large number of patches in a given image to achieve accuracy and confidence gain at the majority voting stage of the proposed system (last block in Figure 1). For the results presented in this paper, we have used $N_p = 128$ as the images in the dataset were quite small. For larger images, a higher value of $N_p$ might be more appropriate to reduce misclassification of each of these patches.

The second step of the proposed system is to read raw DCT coefficients corresponding to each of these patches, directly from the JPEG bitstream. This will avoid introduction of any unwanted noise while performing DCT on the pixel values and will require specialized program as the commonly available image reading program such as imread in Matlab do not provide a way to directly access raw DCT coefficients of an image. We used pysteg [36] for this purpose, which gives raw DCT coefficients $\{D_1, D_2, \ldots, D_f\}$ corresponding to patches $\{P_1, P_2, \ldots, P_f\}$. Given a patch $P_g$, having DCT coefficients $D_g$, final feature vector $H_g$ is obtained by concatenating the histograms of different selective sub-bands. Since the effect of quantization is independent of sign of the DCT coefficients, thus we have taken absolute values of raw DCT coefficients before constructing their histograms. Further, in contrast with some of the existing works, we have used different number of bins in calculating histograms corresponding to different DCT sub-bands. The number of bins is selected based on the position of DCT frequency in zig-zag ordering (value of $k_1 + k_2$ in Figure 2) and lesser number of bins are used for higher values of $k_1 + k_2$. Further, lesser number of bins are allocated to DCT subbands corresponding to chrominance components as compared to those correspond-ing to luminance components (Table 1). Let $B_{Lu}$, and $B_{Cr}$ denote the number of bins in luminance and chrominance channel respectively. For example, for estimating features corresponding to DC coefficients ($k_1 = 1, k_2 = 1$) of luminance channel ($B_{Lu}(k_1 + k_2) = B_{Lu}(2) = 170$), 170 bins are used, number of coefficients with values 1 to 170 are counted (the number of coefficients with 0 value are not used as they are generally very large and do not give any information about quantization step) and normalized by total count of such values. Values above these ranges are neglected as they are occur very rarely.

Let $B$ be the minimum number of bins that must be present in the histogram of a DC subband, for it to be considered in generating the final feature vector. Then, dimensionality $d$ of the final feature vector $H_g$, is given by the following equation:

$$d = \sum_{k_1=1}^{8} \sum_{k_2=1}^{8} \mathbf{u}(9 - k_1 - k_2) \mathbf{u}(B_{Lu}(k_1 + k_2) - B) B_{Lu}(k_1 + k_2)$$
$$+ 2 \sum_{k_1=1}^{8} \sum_{k_2=1}^{8} \mathbf{u}(9 - k_1 - k_2) \mathbf{u}(B_{Cr}(k_1 + k_2) - B) B_{Cr}(k_1 + k_2),$$

where $\mathbf{u}(.)$ is the discrete domain unit step function. In all the experiments reported in this paper, except those analyzing effect of number of DCT subbands (Section 4.5), only those DCT subbands are used which have at least 50 bins in their histogram ($B = 50$) (Table 1), this results in selecting 21 DCT subbands from luminance channel and 3 subbands from each of the chrominance channels and a feature vector $H_g$ of dimensionality d=2230 ($H_g \in \mathbb{R}^{2230 \times 1}$).

Table 1: Variation of Number of Bins in Histograms of Different DCT Subbands

|  | Number of Bins for | |
| --- | --- | --- |
|  | Luminance | Chrominance |
| $k_1 + k_2$ | $B_{Lu}(k_1 + k_2)$ | $B_{Cr}(k_1 + k_2)$ |
| 2 | 170 | 100 |
| 3 | 160 | 50 |
| 4 | 110 | 30 |
| 5 | 90 | 20 |
| 6 | 70 | 10 |
| 7 | 50 | 10 |
| 8 | 45 | 10 |
| 9 | 25 | 10 |

*3.2. Background of Convolution Neural Network*

The functioning of an artificial neural network shows the superficial analogy to the biological neural network, as ANN learns to do a task with the help of the connections between several neurons which are organized into layers. Here, we refer to this analogy as superficial because the biological neurons are much more complex with various types and functions, whilst the neurons in ANN are simple nodes which perform the linear function of separating the training data. To add the nonlinearity in the model, the output of neurons is passed through a nonlinear activation function like sigmoid or tanh. Although the concept of ANN was introduced as early as 1943 [37], during the 70s the interest in ANN was subsided given the complexity in training ANN, the requirement of high processing power, large datasets, etc. However, the backpropagation algorithm, the age of big data, better initialization techniques, advent of GPUs, etc. renewed the interest in ANN and the state of the art performance improved with the help of neural networks. The throughput based GPU design [38] made it possible to converge larger and deeper networks faster than once thought. Parallelization in GPUs also made it possible to test for the various setting of hyper-parameters viz. number of layers, the number of neurons per layer, learning rate, etc. These neural networks with a large number of layers were reintroduced as deep neural networks creating a new branch commonly known as deep learning. The variation of ANN, a convolution neural network (CNN) is designed in a way that maintains the topological structure of the input and weights are shared across the layer. In general, there are mainly three types of layers in a CNN viz. convolution layer, pooling layer, and fully connected layer.

The **convolution layer** consists of several filters each of size, say $m \times n \times C$, where, $m \times n$ is the size of each filter and $C$ is the number of channel of input data. Size of the filter and number of filters are hyper-parameters and architectural choice of the network. Let the filter be denoted as $F$. Given an input $I$ of size $W \times H \times C$, convolutional layer will give the output $G$ of size $W' \times H' \times C'$, where $C'$ is the number of filters of size $m \times n$ in the convolutional layer. In summary, convolutional layer does the following operation [39];

$$I \in \mathbb{R}^{W \times H \times C} \xrightarrow{F \in \mathbb{R}^{m \times n \times C} \ (C' \text{ filters})} G \in \mathbb{R}^{W' \times H' \times C'}.$$

where,

$$G^{c'}(x, y) = b^{c'} + \sum_{s=-(m-1)/2}^{(m-1)/2} \sum_{t=-(n-1)/2}^{(n-1)/2} \sum_{c=1}^{C} F_c^{c'}(s, t) I_c(x+s, y+t),$$

$$\forall c' = 1, 2 \ldots, C' \ \ \forall x = 1, 2 \ldots, M' \ \ \forall y = 1, 2 \ldots, N'$$

$F_c^{c'}$ : $c^{th}$ channel of $F$ for the $c'^{th}$ channel of the output G

$b^{c'}$ : Bias for $c'^{th}$ channel of the output G

$I_c$ : $c^{th}$ channel of input $I$

$G^{c'}$ : $c'^{th}$ channel of output $G$.

A typical CNN architecture has many such convolutional layers and each layer can have different number of filters with different width and height. Non-linearity of model can be taken care with help of activation layer, where activation function such as sigmoid, tanh, Rectified Linear Unit(ReLU) [32] is used. After convolution each element of the output volume $G$ is passed through a non-linear activation function. Rectified Linear Unit(ReLU) is most commonly used non linear activation function. For a given scalar $z$, ReLU activation function $\sigma(z)$ is defined as:

$$\sigma(z) = max(0, z). \tag{1}$$

**Pooling layers** are used to downsample the input as it progresses through the network. Max-pooling is the popular choice of pooling layer where a small window of size $s \times s$ is scanned through the input which selects the pixel with a maximum value in the window. Alternative choices of pooling layers are average-pooling, min-pooling, etc. These pooling layers are optional layers and adding them is again an architectural choice.

In CNN architecture, generally at the end stages of convolution blocks, the output is passed through a **fully connected layer**(FC). The fully connected layer performs the same element-wise dot product as in standard ANN. And at the very end of CNN, there is a softmax layer.

A CNN learns the filter weights and biases of the network by giving the input as labeled data, computing the cost value, and then using the optimization algorithm to update the filter weights by minimizing the cost. Most of the time regularization is also used to reduce the problem of over-fitting the data while minimizing the cost function. There are many regularization algorithms such as L2 regularization, L1 regularization, and Dropout [40] that are used in practice

As the network gets deeper, the small changes in starting layer parameters get amplified and it can be problematic when trying to converge a deep network. One possible solution proposed by Ioffe and Szegedy [41] normalizing each mini-batch of the data before passing it to the next layer using its mean and variance. Adding the batch-normalization(BN) layers has become a standard practice and it alleviates the strong dependence on the weight initialization.

### 3.3. Proposed CNN Architecture

Figure 3 shows the deep CNN architecture proposed in this paper to address the issue of multiple compression detection. In all the experiments reported in this paper, this deep CNN architecture is used for training and testing purpose. Only exception is the results presented in Section 4.4, which deal with optimizing the depth of proposed CNN architecture. Input to the proposed CNN is the 1-D vector obtained in feature extraction step (Section 3.1). Hyper-parameters of the proposed CNN architecture are described in Table 2. It has a total of four convolution layer (Conv1D), four pooling layer (MaxPooling1D), two fully connected layers(FC) and a softmax layer. Weights of all the convolutional layers are initialized with 'He normal initializer' [42]. And bias vectors for each convolutional layer is initialized with zero vector. Batch Normalization(BN) [41]is used before every nonlinear activation function. Rectified Linear Unit (ReLU) is used as the activation function. Max Pooling is performed after each ReLU activations. The dropout rate is set to be 0.1 which refers to setting the weights of a neuron to zero at each update with the probability of 0.1. Batch size of 32 samples is chosen for weight and bias updates. Batch size is also a hyper-parameter of the network. This arrangement of the layers, results in close to 19 million learnable parameters of the network. Adam, an adaptive learning rate optimization algorithm is used for finding the learnable network parameters by optimizing the cost function. Learning rate is initialized with 0.0001, and the value of learning rate is reduced to 0.1 times after every ten epochs. Exponential decay rates $\beta_1$ and $\beta_2$ are chosen to be 0.9 and 0.999 respectively [43] for moment estimation. CNN is trained for fifteen number of epochs and the best model which gives lowest validation error in these 15 epochs is chosen as the final model for evaluating the results on the test images. All of our experiments are performed on a NVIDIA GeForce GTX 1080 GPU with 8 GB memory.

## 4. Experimental Result

### 4.1. The Database

For the experimental validation UCID [26] dataset consisting of 1338 uncompressed color images in TIFF (Tagged Image File Format) having resolution of 512×384 or 384×512 is used. For the experimental result, data is split into a training set, validation set, and test set.

The nomenclature used for the images compressed at once, twice or N number of times are termed as images from class $C_1$, $C_2$, or $C_N$ respectively. Dataset generation procedure is as follows: For the generation of images of class $C_1$, say, $p$ number of uncompressed images from the UCID dataset are compressed at the quality factor, say $QF_N$. Now for creating images for the class $C_2$, a chain $(QF_{N-1}, QF_N)$ of $r$ unique quality factor is generated using Equation 2. All the $p$ images are compressed with these $r$ unique chain of quality factor based on the constraint described in Equation 2. The similar procedure is applied for creating the images of $C_3$, $C_4$, and $C_N$. This will result in $C_1$ class having $p$ number of images, and all other classes will have $p*r$ number of images. To balance the number of images for each class, $p$ images of $C_1$ class are repeated $r$ times. Now each class has $p*r$ number of images. All the images of all the classes will have last quality factor of $QF_N$. The default value of $r$ is chosen to be 10.

In this paper number of training and testing images are reported in terms $p$. For example number of training or testing image 200 means, each class has 2000 images as the value of $r$ is fixed to 10. Images from each class are split in to patch size of 128× 128. 80% of the patches from each class are used for training and remaining 20% is used for validation purpose. For all the experiments, testing is done with the patches from different 500 images from the dataset.

For generating the images compressed up to N times [29], a unique chain of compression is being generated as follows: the Quality factor of $i^{th}$ compression stage is randomly chosen from all possible values of $QF_i$, which is described in Equation 2.

$$\begin{aligned} QF_i &\in \left(QF_{i+1}^l \cup QF_{i+1}^u\right) \cap ([Q_{min} : Q_{max}]) \\ \text{where} \quad QF_{i+1}^l &= [QF_{i+1} - dq_{max} : QF_{i+1} - dq_{min}], \\ QF_{i+1}^u &= [QF_{i+1} + dq_{min} : QF_{i+1} + dq_{max}]. \end{aligned} \quad (2)$$

Where $Q_{min}$ and $Q_{max}$ are the global minimum and maximum value of quality factor, that is fixed to be 60 and 95 respectively. $dq_{min}$ and $dq_{max}$ are the quantities which make sure that value of the quality factor($QF_i$) is not too close and not too far from the ($QF_{i+1}$). Value of $dq_{min}$ and $dq_{max}$ are chosen to be 6 and 12. These empirical numbers are adapted from [29].

The default values of number of training images $p$, number of test images, last quality factor $QF_N$, and number of compressions are chosen to be 200, 500, 80 and 4 unless stated otherwise.

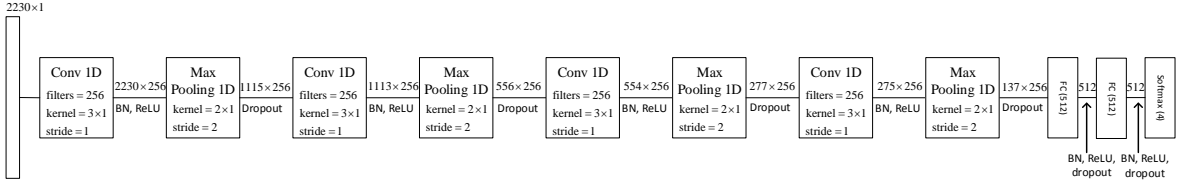

Figure 3: The Proposed CNN Architecture

Table 2: CNN Architecture Details

| Layer | Input Size | Filter Size | Stride | Number of Filters | Output Size | Number of Parameters |
|---|---|---|---|---|---|---|
| Conv1D-1 | 2230 × 1 | 3 × 1 | 1 | 256 | 2230 × 256 | 1024 |
| BN-1, ReLU-1 | 2230 × 256 | - | - | - | 2230 × 256 | 1024, 0 |
| MaxPooling1D-1 | 2230 × 256 | 2 × 1 | 2 | - | 1115 × 256 | 0 |
| Dropout-1 | 1115 × 256 | - | - | - | 1115 × 256 | 0 |
| Conv1D-2 | 1115 × 256 | 3 × 1 | 1 | 256 | 1113 × 256 | 196864 |
| BN-2, ReLU-2 | 1113 × 256 | - | - | - | 1113 × 256 | 1024, 0 |
| MaxPooling1D-2 | 1113 × 256 | 2 × 1 | 2 | - | 556 × 256 | 0 |
| Dropout-2 | 556 × 256 | - | - | - | 556 × 256 | 0 |
| Conv1D-3 | 556 × 256 | 3 × 1 | 1 | 256 | 554 × 256 | 196864 |
| BN-3, ReLU-3 | 554 × 256 | - | - | - | 554 × 256 | 1024, 0 |
| MaxPooling1D-3 | 554 × 256 | 2 × 1 | 2 | - | 277 × 256 | 0 |
| Dropout-3 | 277 × 256 | - | - | - | 277 × 256 | 0 |
| Conv1D-4 | 277 × 256 | 3 × 1 | 1 | 256 | 275 × 256 | 196864 |
| BN-4, ReLU-4 | 275 × 256 | - | - | - | 275 × 256 | 1024, 0 |
| MaxPooling1D-4 | 275 × 256 | 2 × 1 | 2 | - | 137 × 256 | 0 |
| Dropout-4 | 137 × 256 | - | - | - | 137 × 256 | 0 |
| FC-1 (512 neurons) | 137 × 256 (flatten) | - | - | - | 512 | 17957376 |
| BN-5, ReLU-5 | 512 | - | - | - | 512 | 2048,0 |
| Dropout-5 | 512 | - | - | - | 512 | 0 |
| FC-2 (512 neurons) | 512 | - | - | - | 512 | 262656 |
| BN-6, ReLU-6 | 512 | - | - | - | 512 | 2048, 0 |
| Dropout-6 | 512 | - | - | - | 512 | 0 |
| Softmax (N neurons) | 512 | - | - | - | N | 512*N+N |

*4.2. Effect of Number of Training Images*

As described in the previous section, we have chosen 200 images as our default number of training images, in which 80% of the patches are used for training purpose and remaining 20% patches for validation purpose. And testing set contains patches from 500 different images. In this section, we have shown experimentally why choosing patches from 200 images is reasonable. As images from UCID dataset have the images of resolution 512 × 384 or 384 × 512. We will get 12 patches of size 128 × 128 for each image. Based on the description in Section 4.1, each class will have 2000 number of images, hence 24000 patches for each class. Now 80% of these patches(22,800) are used for training and remaining 20% is used for validation purpose.

Table 3 indicates the average test accuracy with 500 test images. The number of training images is varied from 100 to 500 in the steps of 100. For each training image set such as 100, CNN is trained for five times and average test accuracies reported in the Table 3 is the mean of accuracies obtained from each of the five models. From this Table 3 we can conclude that increasing the number of train images, average accuracy goes towards saturation, but training time of CNN per epoch increases due to large training images. Keeping the computation time into consideration, we have chosen 200 images as our default choice for training the CNN with the expanse of some accuracy.

Note that 'average time per epoch' will depend on the specifications of GPU used for training the CNN architecture, but the trend in average time per epoch will remain the same with all kind of GPU's.

Table 3: Training Size Variation

| Number of training images | Average accuracy(%) | Standard deviation(%) | Average time per epoch (in minutes) |
|---|---|---|---|
| 100 | 96.87 | 0.15 | 1.87 |
| 200 | 97.79 | 0.19 | 3.75 |
| 300 | 98.08 | 0.13 | 5.60 |
| 400 | 98.11 | 0.25 | 7.47 |
| 500 | 98.26 | 0.15 | 9.60 |

## 4.3. Effect of Number of Epochs

We wanted to optimize the number of epochs for training the CNN architecture. To accomplish this, we fixed the number of training images to 200 and CNN architecture to that shown in Figure 3. Figure 4 shows the average patch level train and validation accuracy, while Figure 5 shows average train and validation loss with standard deviation. These average values are the result of training the same CNN model for six times. The motivation for performing this kind of experiment was to find out the number of epochs sufficient to train the model. And we wanted to see if we can evaluate the result with single trained CNN instead of using an ensemble of CNNs. We observed that average accuracy tends towards saturation and standard deviation of average validation accuracy reduces, that conclude that fifteen number of epochs is sufficient to train the model. And testing with single trained model results in a stable result with the standard deviation of 0.19% (row 3 of Table 3).

Note that all the results, reported hereafter, are evaluated with the single trained CNN models.

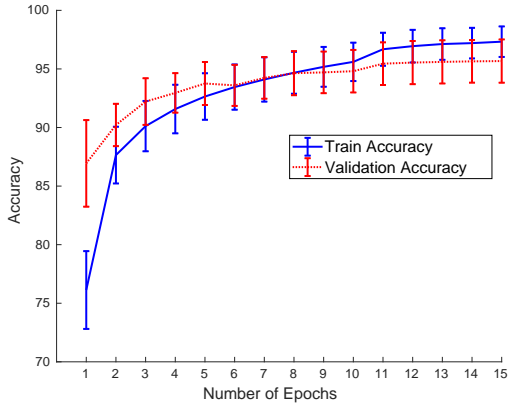

Figure 4: Effect of Number of Epochs on Model Accuracy

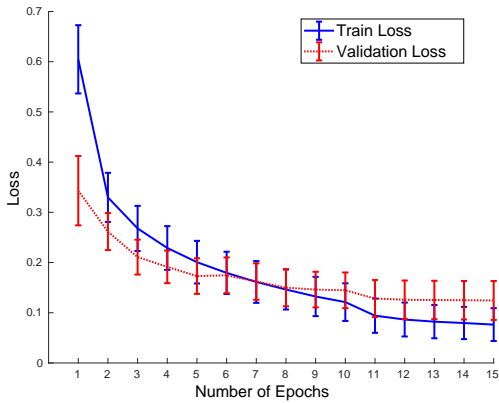

Figure 5: Effect of Number of Epochs on Model Loss

## 4.4. Effect of Depth of the CNN

In this experiment number of training images and number of epochs are fixed, but the depth of the CNN architecture is varied. In proposed CNN shown in Figure 3, there is four convolutional layer. After each convolutional layer, there is batch normalization (BN), Rectified Linear Unit (ReLU) activation, Maxpooling, and Dropout. In the Table 4, different CNN architectures which are named as CNN1, CNN2, to CNN6 are shown. CNN2 is our proposed architecture. Each architecture has same convolutional and pooling layer as shown Figure 3 and described in Table 2. For the compactness purpose Batch Normalization, ReLU and Dropout have not been shown in the Table 4. And the accuracies reported are on the set of default 500 number of test images. From the Table 4, it is evident that CNN2 with close to 19 million parameters and average test accuracy of 98.09% is a good model to choose for the evaluation of results.

## 4.5. Effect of Number of Sub-bands for Histogram Calculation

After finding the best choices of the number of training images, the number of epochs, and final CNN architecture, this section analyzes the effect of increasing the number of sub-bands for histogram calculation. As we increase the number of sub-bands for the histogram calculation, the performance of the algorithm increases slightly (Table 5). Again slight improvement in the performance of the algorithm is expected as even though the majority of the quantized JPEG coefficients are zero at higher sub-bands, still there are some non-zero coefficients. But considering computation time (Table 5), we have used first 21 sub-bands from luminance channel and three sub-bands from each of the two chrominance channels, as our default sub-bands for histogram calculation, which results in feature dimensionality of $2230 \times 1$. This default choice of the sub-bands is compared with when first 36 sub-bands from each of the luminance and chrominance channels are selected to result in the dimensionality of $3605 \times 1$.

## 4.6. Analysis with Different $QF_N$ and Number of Compression Stages

After fixing the number of training images, test images, and final CNN architecture, we evaluated the performance of the algorithm with the image undergone up to four and five compression stages. Extensive experiments are performed to evaluate the performance of the proposed algorithm. We have tested the algorithm with last quality factor($QF_N = \{75, 80, 85\}$) as these are the frequently used last quality factors for re-saving the image in JPEG format. Based on the quality factor chosen from the Equation 2 for previous compressions, chain of quality factor for different stages of image compression will have all kind of quality factor such as $QF_i > QF_{i+1}$ and $QF_i < QF_{i+1}$(while satisfying assumption of Equation 2). Confusion matrices for four number of JPEG compression with $QF_N = 75, QF_N = 80$ and $QF_N = 85$ are shown in Table 6, 7 and 8 respectively. Experimental results for the images undergone upto five compression stages with the values of $QF_N = 75$, $QF_N = 80$ and

Table 4: Effect of depth of proposed CNN* on test accuracy(*BN, ReLU and Dropout is omitted for compactness purpose)

| | CNN1 | CNN2 | CNN3 | CNN4 | CNN5 | CNN6 |
|---|---|---|---|---|---|---|
| | Conv1D-1 Maxpooling1D-1 | Conv1D-1 Maxpooling1D-1 | Conv1D-1 Maxpooling1D-1 | Conv1D-1 Maxpooling1D-1 | Conv1D-1 Maxpooling1D-1 | Conv1D-1 Maxpooling1D-1 |
| | Conv1D-2 Maxpooling1D-2 | Conv1D-2 Maxpooling1D-2 | Conv1D-2 Maxpooling1D-2 | Conv1D-2 Maxpooling1D-2 | Conv1D-2 Maxpooling1D-2 | Conv1D-2 Maxpooling1D-2 |
| | Conv1D-3 Maxpooling1D-3 | Conv1D-3 Maxpooling1D-3 | Conv1D-3 Maxpooling1D-3 | Conv1D-3 Maxpooling1D-3 | Conv1D-3 Maxpooling1D-3 | Conv1D-3 Maxpooling1D-3 |
| | FC-1 | Conv1D-4 Maxpooling1D-4 | Conv1D-4 Maxpooling1D-4 | Conv1D-4 Maxpooling1D-4 | Conv1D-4 Maxpooling1D-4 | Conv1D-4 Maxpooling1D-4 |
| | FC-2 | FC-1 | Conv1D-5 Maxpooling1D-5 | Conv1D-5 Maxpooling1D-5 | Conv1D-5 Maxpooling1D-5 | Conv1D-5 Maxpooling1D-5 |
| | Softmax | FC-2 | FC-1 | Conv1D-6 Maxpooling1D-6 | Conv1D-6 Maxpooling1D-6 | Conv1D-6 Maxpooling1D-6 |
| | | Softmax | FC-2 | FC-1 | Conv1D-7 Maxpooling1D-7 | Conv1D-7 Maxpooling1D-7 |
| | | | Softmax | FC-2 | FC-1 | Conv1D-8 Maxpooling1D-8 |
| | | | | Softmax | FC-2 | FC-1 |
| | | | | | Softmax | FC-2 |
| | | | | | | Softmax |
| Test Accuracy (%) | 97.56 | 97.79 | 97.54 | 97.19 | 95.94 | 94.58 |
| # Parameters | 36,974,084 | 18,821,892 | 9,844,740 | 5,455,108 | 3,424,772 | 2,443,012 |
| Time per epoch (in Min.) | 4.3 | 3.75 | 3.57 | 3.52 | 3.34 | 3.36 |

Table 5: Variation in Number of Sub-bands for Histogram Calculation

| # Sub-bands | Average accuracy(%) | Average time per epoch(in minutes) |
|---|---|---|
| (21,3,3) | 97.79 | 3.75 |
| (36,36,36) | 98.89 | 6.067 |

$QF_N = 85$ are shown in Table 9, Table 10 and Table 11 respectively. We can observe the trend that as we increase the value of last quality factor $QF_N$ in both the cases when the images are compressed up to four times and five times, performance of the algorithm increases, due to increase in non-zero quantized coefficients in each image. With these confusion matrices given in this Section, we show that our method can detect reliably up to five compression stages.

Table 6: Confusion matrix for $QF_N = 75$

| | $C_1$ | $C_2$ | $C_3$ | $C_4$ |
|---|---|---|---|---|
| $C_1$ | 96.60 | 0.00 | 3.40 | 0.00 |
| $C_2$ | 1.14 | 97.38 | 0.80 | 0.68 |
| $C_3$ | 19.30 | 1.22 | 79.30 | 0.18 |
| $C_4$ | 0.24 | 2.52 | 0.14 | 97.10 |

Table 7: Confusion matrix for $QF_N = 80$

| | $C_1$ | $C_2$ | $C_3$ | $C_4$ |
|---|---|---|---|---|
| $C_1$ | 96.60 | 1.40 | 2.00 | 0.00 |
| $C_2$ | 0.16 | 99.42 | 0.40 | 0.02 |
| $C_3$ | 4.88 | 0.14 | 94.96 | 0.02 |
| $C_4$ | 0.00 | 0.02 | 0.04 | 99.94 |

Table 8: Confusion matrix for $QF_N = 85$

| | $C_1$ | $C_2$ | $C_3$ | $C_4$ |
|---|---|---|---|---|
| $C_1$ | 99.40 | 0.60 | 0.00 | 0.00 |
| $C_2$ | 0.36 | 99.30 | 0.14 | 0.20 |
| $C_3$ | 1.56 | 0.16 | 98.28 | 0.00 |
| $C_4$ | 0.00 | 0.36 | 0.02 | 99.62 |

Table 9: Confusion matrix for $QF_N = 75$

| | $C_1$ | $C_2$ | $C_3$ | $C_4$ | $C_5$ |
|---|---|---|---|---|---|
| $C_1$ | 92.40 | 0.20 | 7.40 | 0.00 | 0.00 |
| $C_2$ | 1.02 | 96.66 | 1.16 | 1.14 | 0.02 |
| $C_3$ | 15.72 | 0.74 | 83.48 | 0.06 | 0.00 |
| $C_4$ | 0.26 | 1.38 | 0.16 | 98.20 | 0.00 |
| $C_5$ | 0.02 | 0.00 | 0.06 | 0.00 | 99.92 |

Table 10: Confusion matrix for $QF_N = 80$

| | $C_1$ | $C_2$ | $C_3$ | $C_4$ | $C_5$ |
|---|---|---|---|---|---|
| $C_1$ | 97.40 | 1.20 | 1.40 | 0.00 | 0.00 |
| $C_2$ | 0.32 | 99.32 | 0.34 | 0.02 | 0.00 |
| $C_3$ | 5.24 | 0.14 | 94.58 | 0.02 | 0.02 |
| $C_4$ | 0.00 | 0.04 | 0.04 | 99.92 | 0.00 |
| $C_5$ | 0.00 | 0.02 | 0.08 | 0.04 | 99.86 |

*4.7. Comparison with Existing Work*

The performance of the proposed method is compared with the method proposed in [29] with the value of $QF_N = 80$. In [29], authors, have used 100 images from UCID dataset for training the SVM classifier and 200 images for testing purpose. Note that the confusion matrix reported here is directly taken from the author's paper. This Table 12 corresponds to Table

Table 11: Confusion matrix for $QF_N = 85$

|       | $C_1$ | $C_2$ | $C_3$ | $C_4$ | $C_5$ |
|-------|-------|-------|-------|-------|-------|
| $C_1$ | 99.20 | 0.40  | 0.40  | 0.00  | 0.00  |
| $C_2$ | 0.44  | 98.82 | 0.00  | 0.74  | 0.00  |
| $C_3$ | 1.34  | 0.14  | 98.50 | 0.02  | 0.00  |
| $C_4$ | 0.00  | 0.10  | 0.00  | 99.90 | 0.00  |
| $C_5$ | 0.00  | 0.02  | 0.10  | 0.02  | 99.86 |

2(a) of the author's paper in [29]. For the comparison purpose, we have also trained the model with 100 images and tested with 200 images. Note that one to one comparison of these two Tables 12 and 13 is not possible because of possibly different 200 images form UCID dataset used for training the SVM in [29] and CNN in our case, and also because of the random chain of compression used in both the cases. But the comparison with average accuracy can provide some better insight as our methods give 97.45%, while method in [29] gives the average accuracy of 87.93%. Hence our method performs significantly better with average accuracy gain of nearly 10%.

Table 12: Confusion matrix for $QF_N = 80$ [29](adapted)

|       | $C_1$  | $C_2$ | $C_3$ | $C_4$ |
|-------|--------|-------|-------|-------|
| $C_1$ | 100.00 | 0.00  | 0.00  | 0.00  |
| $C_2$ | 2.09   | 94.18 | 1.52  | 2.21  |
| $C_3$ | 0.20   | 1.52  | 71.23 | 27.05 |
| $C_4$ | 0.00   | 0.92  | 12.75 | 86.32 |

Table 13: Confusion matrix for $QF_N = 80$ (100 Training and 200 Testing Images)

|       | $C_1$ | $C_2$ | $C_3$ | $C_4$ |
|-------|-------|-------|-------|-------|
| $C_1$ | 97.50 | 0.00  | 2.50  | 0.00  |
| $C_2$ | 0.25  | 99.35 | 0.35  | 0.05  |
| $C_3$ | 6.70  | 0.10  | 93.20 | 0.00  |
| $C_4$ | 0.00  | 0.00  | 0.25  | 99.75 |

### 4.8. Patch Level Compression Detection

Figure 6 shows the results on $128 \times 128$ patches of four test images from each class, out of the 500 test images compressed up to four times with $QF_N = 80$. On $128 \times 128$ patches of all the 500 test images, the performance of the algorithm in terms of average accuracy is 91%.

Color coding of the patches is as following: Misclassified patches from the class $C_1$, $C_2$, $C_3$, and $C_4$ are shown in white, red, green and blue respectively while correctly classified patches are shown in their original color. One important thing to note is that majority of the misclassified patches are from the non-textured regions of the images. This observation is valid for most of the images in all the four classes $C_1$, $C_2$, $C_3$, and $C_4$ in the test set of 500 images. Here we have shown only four images from each of the classes.

## 5. Conclusion

In this paper, we have proposed first of its kind system for multiple JPEG compression classification using DCT domain deep CNN. We have designed appropriate pre-processing stage before feeding the image data to CNN. The proposed pre-processing stage directly utilizes JPEG bit stream and reduces the content dependent nature of the data fed to the CNN. Existing systems utilizing CNN had demonstrated their applicability on double JPEG compression detection only. The proposed method outperformed handcrafted features-based existing system for multiple JPEG compression detection on the experimental scenarios reported in the literature. Further, the proposed system is capable of efficiently handling a larger number of re-compression cycles then the existing systems. Experimental results show that its performance does not deteriorate even up to five compression cycles and thus its promising for the scenarios involving an even larger number of compression cycles. Future work will include extending the proposed method for forgery localization as it also gives excellent performance in classifying patches of size $128 \times 128$.

## Acknowledgment

This material is based upon work partially supported by a grant from the Department of Science and Technology (DST), New Delhi, India, under Award Number ECR/2015/000583 and Indian Institute of Technology Gandhinagar internal research grant IP/IITGN/EE/NK/201516-06. Any opinions, findings, and conclusions or recommendations expressed in this material are those of the author(s) and do not necessarily reflect the views of the funding agencies. Address all correspondence to Nitin Khanna at nitinkhanna@iitgn.ac.in.

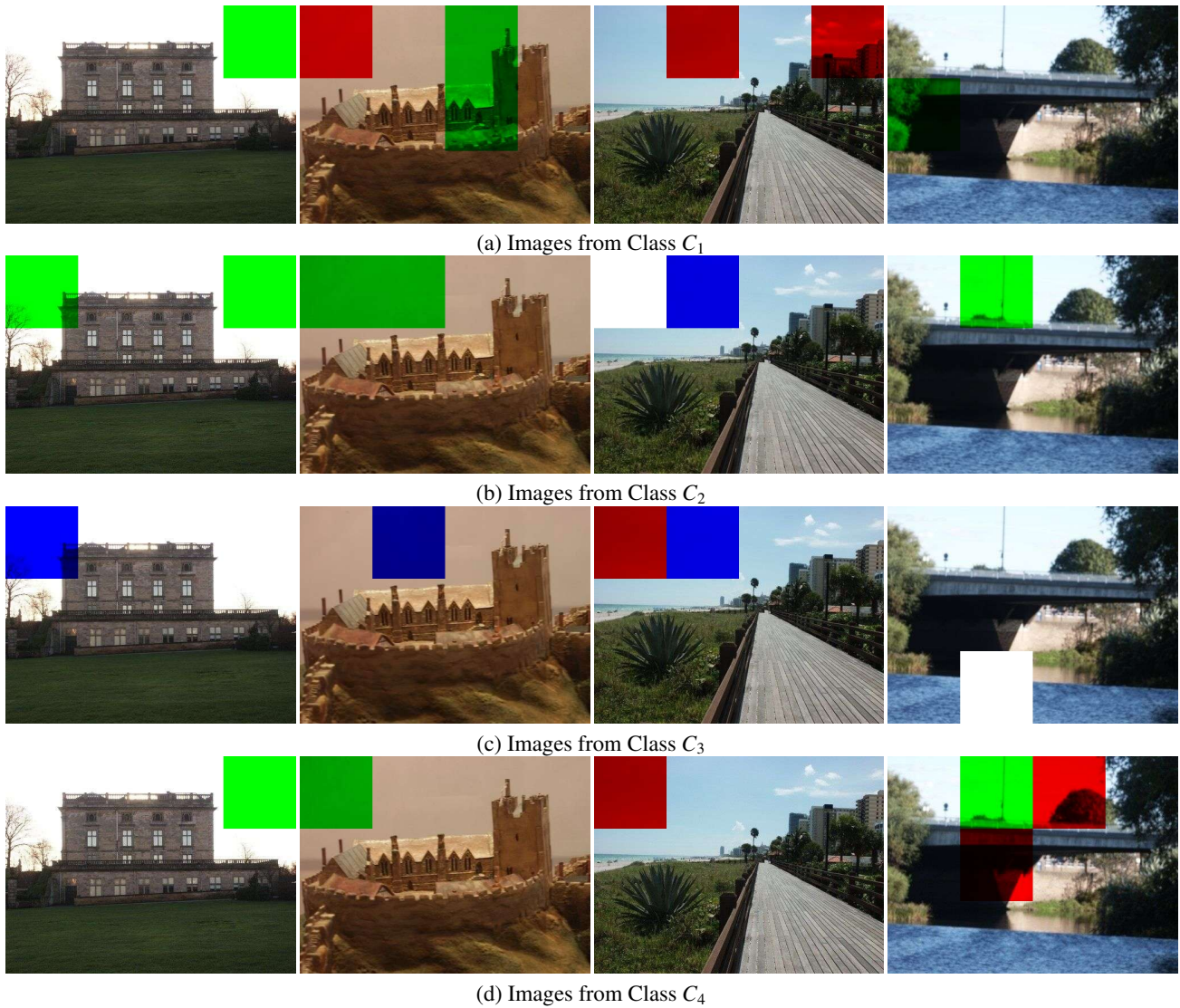

Figure 6: Patch Level Compression Detection with $QF_N = 80$ (Correctly predicted patches are in their original color while misclassified patches from the class $C_1$, $C_2$, $C_3$ and $C_4$ are shown in white, red, green and blue, respectively)